\documentclass[sigconf]{acmart}
\usepackage{enumitem}
\usepackage{subfigure}
\usepackage{amsfonts}
\usepackage{amsmath}
\usepackage{makecell}
\usepackage{multirow}
\usepackage{balance}
\usepackage[ruled]{algorithm2e} 
\AtBeginDocument{%
  \providecommand\BibTeX{{%
    \normalfont B\kern-0.5em{\scshape i\kern-0.25em b}\kern-0.8em\TeX}}}

\copyrightyear{2023}
\acmYear{2023}
\setcopyright{acmlicensed}\acmConference[RecSys '23]{Seventeenth ACM Conference on Recommender Systems}{September 18--22, 2023}{Singapore, Singapore}
\acmBooktitle{Seventeenth ACM Conference on Recommender Systems (RecSys '23), September 18--22, 2023, Singapore, Singapore}
\acmPrice{15.00}
\acmDOI{XXXXXXX.XXXXXXX}

\begin{document}

\title{Augmented Negative Sampling for Collaborative Filtering}
\author{Yuhan Zhao}
\affiliation{
  \institution{Harbin Engineering University}
  \city{Harbin}
  \country{China}
}
\email{asa9ao@hrbeu.edu.cn}

\author{Rui Chen}
\authornote{Corresponding authors.}
\affiliation{
  \institution{Harbin Engineering University}
  \city{Harbin}
  \country{China}
}
\email{ruichen@hrbeu.edu.cn}

\author{Riwei Lai}
\affiliation{
  \institution{Harbin Engineering University}
  \city{Harbin}
  \country{China}
}
\email{lai@hrbeu.edu.cn}

\author{Qilong Han}
\authornotemark[1]
\affiliation{
  \institution{Harbin Engineering University}
  \city{Harbin}
  \country{China}
}
\email{hanqilong@hrbeu.edu.cn}

\author{Hongtao Song}
\affiliation{
  \institution{Harbin Engineering University}
  \city{Harbin}
  \country{China}
}
\email{songhongtao@hrbeu.edu.cn}

\author{Li Chen}
\affiliation{
  \institution{Hong Kong Baptist University}
  \city{Hong Kong}
  \country{China}
}
\email{lichen@comp.hkbu.edu.hk}

\renewcommand{\shortauthors}{Yuhan Zhao, Rui Chen, Riwei Lai, Qilong Han, Hongtao Song \&  Li Chen}

\begin{abstract}
Negative sampling is essential for implicit-feedback-based collaborative filtering, which is used to constitute negative signals from massive unlabeled data to guide supervised learning. The state-of-the-art idea is to utilize hard negative samples that carry more useful information to form a better decision boundary. To balance efficiency and effectiveness, the vast majority of existing methods follow the two-pass approach, in which the first pass samples a fixed number of unobserved items by a simple static distribution and then the second pass selects the final negative items using a more sophisticated negative sampling strategy. However, selecting negative samples from the original items in a dataset is inherently restricted due to the limited available choices, and thus may not be able to contrast positive samples well. In this paper, we confirm this observation via carefully designed experiments and introduce two major limitations of existing solutions: ambiguous trap and information discrimination.

Our response to such limitations is to introduce ``augmented'' negative samples that may not exist in the original dataset. This direction renders a substantial technical challenge because constructing unconstrained negative samples may introduce excessive noise that eventually distorts the decision boundary. To this end, we introduce a novel generic augmented negative sampling (ANS) paradigm and provide a concrete instantiation. First, we disentangle hard and easy factors of negative items. Next, we generate new candidate negative samples by augmenting only the easy factors in a regulated manner: the direction and magnitude of the augmentation are carefully calibrated. Finally, we design an advanced negative sampling strategy to identify the final augmented negative samples, which considers not only the score function used in existing methods but also a new metric called augmentation gain. Extensive experiments on real-world datasets demonstrate that our method significantly outperforms state-of-the-art baselines. Our code is publicly available at \url{https://github.com/Asa9aoTK/ANS-Recbole}.

\end{abstract}



\maketitle

\section{Introduction} 
Collaborative filtering (CF), as an important paradigm of recommender systems, leverages observed user-item interactions to model users' potential preferences~\cite{HDW20,HLZ17,WHW19}. In real-world scenarios, such interactions are normally in the form of implicit feedback (e.g., clicks or purchases), instead of explicit ratings~\cite{WFH21}.Each observed interaction is normally considered a positive sample. As for negative samples, existing methods usually randomly select some uninteracted items. Then a CF model is optimized to give positive samples higher scores than negative ones via, for example, the Bayesian personalized ranking (BPR) loss function~\cite{RFG12}, where a score function (e.g., inner product) is used to measure the similarity between a user and an item.

Recent studies have shown that negative samples have a great impact on model performance~\cite{ZZY22,MZW21,ZCW13}. As the state of the art, \emph{hard} negative sampling strategies whose general idea is to oversample high-score negative items have exhibited promising performance~\cite{ZCW13,DQH19,DQY20,ZZH22,CLJ22}. While selecting a negative item with a high score makes it harder for a model to classify a user, it has the potential to bring more useful information and greater gradients, which are beneficial to model training~\cite{RF14,CLJ22}. Ideally, one would calculate the scores of all uninteracted items to identify the best negative samples. However, its time complexity is prohibitive. To balance efficiency and effectiveness, the two-pass approach has been widely adopted~\cite{BGZ17,ZCW13,DQY20,ZZH22,HDD21}. The first pass samples a fixed number of unobserved items by a static distribution, 
and the second pass then selects the final negative items with a more sophisticated negative sampling method. 

Despite the significant progress made by hard negative sampling, selecting negative samples from the original items in a dataset is inherently restricted due to the limited available choices. Such original items may not be able to contrast positive samples well. Indeed, we design a set of intuitive experiments to show that existing works suffer from two major drawbacks. (1) \emph{Ambiguous trap.} Since the vast majority of unobserved items have very low scores (i.e., they are easy negative samples), randomly sampling a small number of candidate negative items in the first pass is difficult to include useful hard negative samples, which, in turn, substantially limits the efficacy of the second pass. (2) \emph{Information discrimination.} In the second pass, most existing studies overly focus on high-score negative items and largely neglect low-score negative items. We empirically show that such low-score negative items also contain critical, unique information that leads to better model performance.

Our response to such drawbacks is to introduce \emph{``augmented'' negative samples} (i.e., synthetic items) that are more similar to positive items while still being negative. While data augmentation techniques have been proposed in other domains~\cite{KSP20,YYX22,WWF21,XYY21}, it is technically challenging to apply similar ideas to negative sampling for collaborative filtering. This is because all of them fail to carefully regulate and quantify the augmentation needed to approximate positive items while not introducing excessive noise or still being negative. 

To this end, we present a novel generic augmented negative sampling (ANS) paradigm and then provide a concrete instantiation. Our insight is that it is imperative to understand a negative item's hardness from a more fine-granular perspective. We propose to disentangle an item's embedding into hard and easy factors (i.e., a set of dimensions of the embedding vector), where hardness is defined by whether a negative item has similar values to the corresponding user in the given factor. This definition is in line with the definition of hardness in hard negative sampling. Here the key technical challenge originates from the lack of supervision signals. Consequently, we propose two learning tasks that combine contrastive learning (CL)~\cite{ZGC22} and disentanglement methods~\cite{HWN22} to guarantee the credibility of the disentanglement. Since our goal is to create synthetic negative items similar to positive items, we keep the hard factor of a negative item unchanged and focus on augmenting the easy factor by controlling the direction and magnitude of added noise. The augmentation mechanism needs to be carefully designed so that the augmented item will become more similar to the corresponding positive item, but will not cross the decision boundary.
Furthermore, we introduce a new metric called augmentation gain to measure the difference between the scores before and after the augmentation. Our sampling strategy is guided by augmentation gain, which gives low-score items with higher augmentation gain a larger probability of being sampled. In this way, we can effectively mitigate information discrimination, leading to better performance.
We summarize our main contributions as follows:
\begin{itemize}[leftmargin=*]
    \item We design a set of intuitive experiments to reveal two notable limitations of existing hard negative sampling methods, namely ambiguous trap and information discrimination. 
    \item To the best of our knowledge, we are the first to propose to generate negative samples from a fine-granular perspective to improve implicit CF. In particular, we propose a new direction of generating regulated augmentation to address the unique challenges of CF. 
    \item We propose a general paradigm called augmented negative sampling (ANS) that consists of three steps, including disentanglement, augmentation, and sampling. We also present a concrete implementation of ANS, which is not only performant but also efficient. 
    \item We conduct extensive experiments on five real-world datasets to demonstrate that ANS can achieve significant improvements over representative state-of-the-art negative sampling methods.
\end{itemize}

\section{Related Work}

\subsection{Model-Agnostic Negative Sampling}
A common type of negative sampling strategy selects negative samples based on a pre-determined static distribution~\cite{CSS17,WVS19,YDZ20}. Such a strategy is normally efficient since it does not need to be adjusted in the model training process. Random negative sampling (RNS)~\cite{RFG12,WHW19,YYX22,CWH20} is a simple and representative model-agnostic sampling strategy, which selects negative samples from unobserved items according to a uniform distribution. However, the uniform distribution is difficult to guarantee the quality of negative samples. Inspired by the word-frequency-based distribution~\cite{GL16} and node-degree-based distribution~\cite{MSC13} in other domains, an item-popularity-based distribution~\cite{CSS17,BS08} has been introduced. Under this distribution, popular items are more likely to be sampled as negative items, which helps to mitigate the widespread popularity bias issue in recommender systems~\cite{CDW20}. 
Although this kind of strategy is generally efficient, the pre-determined distributions are not customized for the underlying models and not adaptively adjusted during the training process. As a result, their performance is often sub-optimal.

\subsection{Model-Aware Negative Sampling}
These strategies take into consideration some information of the underlying model, denoted by $f$, to guide the sampling process. Given $f$, the probability of sampling an item $i$ is defined as $p(i \mid f) \propto g(f,\mathbf{e}_i)$, where $g(\cdot, \cdot)$ is a sampling function, and $\mathbf{e}_i$ denotes the embedding of $i$. Existing studies focus on choosing different $f$ and/or designing a proper $g(\cdot, \cdot)$ to achieve better performance. The most representative work is hard negative sampling, which defines $g(\cdot, \cdot)$ as a score function. It assigns higher sampling probabilities to the negative items with larger prediction scores~\cite{DQH19,DQY20,HDD21,RF14,ZCW13,ZZH22}.
For example, DNS~\cite{ZCW13} assumes that the high-score items should be more likely to be selected, and thus chooses $g(\cdot, \cdot)$ to be the inner product and $f$ to be user representations. With the goal of mitigating false negative samples, SRNS~\cite{DQY20} further incorporates the information about the last few epochs into $f$ and designs $g(\cdot, \cdot)$ to give false negative samples lower scores. IRGAN~\cite{WYZ17} integrates a generative adversarial network into $g(\cdot, \cdot)$ to determine the probabilities of negative samples through the min-max game. ReinforcedNS~\cite{DQH19} use reinforcement learning into $g(\cdot, \cdot)$.

With well-designed $f$ and $g(\cdot, \cdot)$, we can generally achieve better performance. However, selecting suitable negative items need to compute $g(\cdot, \cdot)$ for all unobserved items, which is extremely time-consuming and prohibitively expensive. Take DNS as an example. Calculating the probability of sampling an item is equivalent to performing softmax on all unobserved samples, which is unacceptable in real-world applications~\cite{CLJ22,WYZ17,PC19}. As a result, most model-aware sampling strategies adopt the two-pass approach or its variants. In this case, $g(\cdot, \cdot)$ is only applied to a small number of candidates sampled in the first pass. While such two-pass-based negative sampling strategies have been the mainstream methods, they exhibit two notable limitations, namely ambiguous trap and information discrimination, which motivates us to propose an augmented negative sampling paradigm. In the next section, we will explain these limitations via a set of intuitive experiments.

\begin{figure*}[t]
\center
\subfigure[Epoch 0]{
\begin{minipage}[t]{0.23\linewidth} 
\centering
\includegraphics[width=\linewidth]{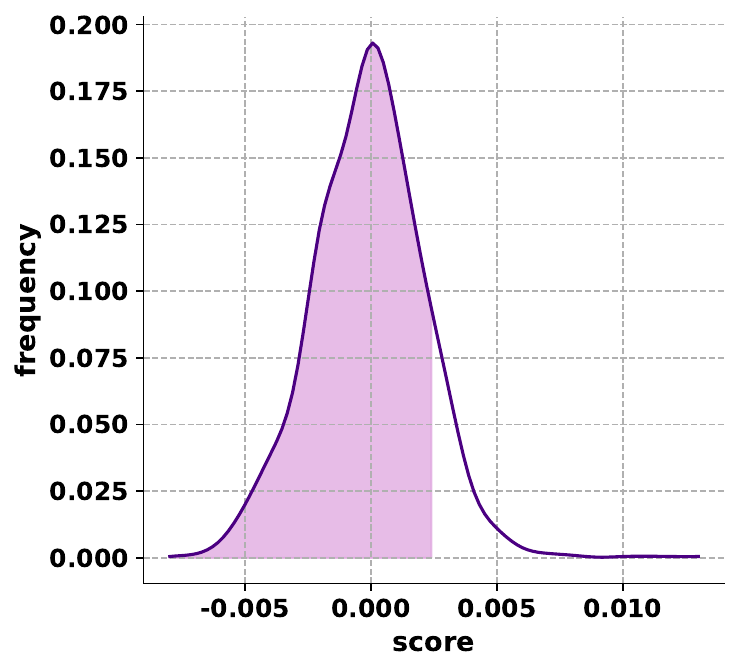}
\label{fig:epoch0}
\end{minipage}
}
\subfigure[Epoch 30]{
\begin{minipage}[t]{0.23\linewidth}
\centering
\includegraphics[width=\linewidth]{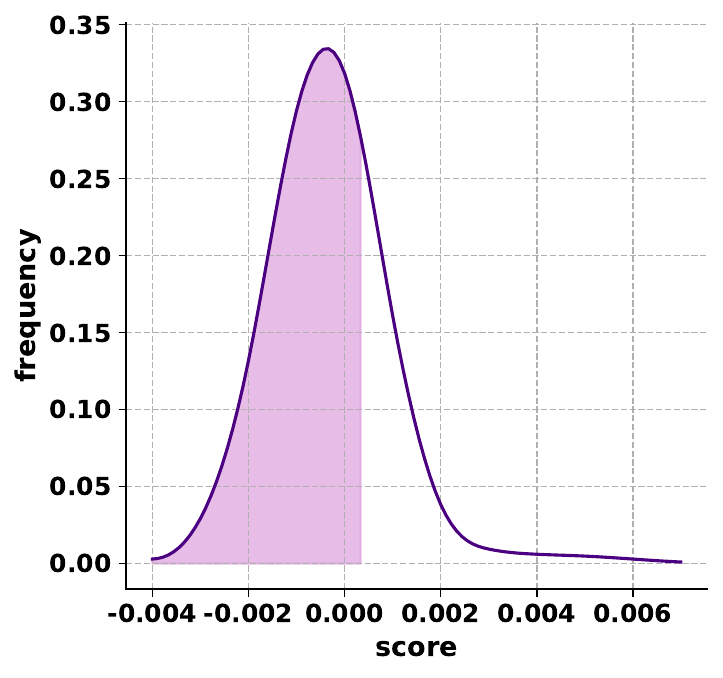}
\label{fig:epoch30}
\end{minipage}
}
\subfigure[Epoch 50]{
\begin{minipage}[t]{0.23\linewidth}
\centering
\includegraphics[width=\linewidth]{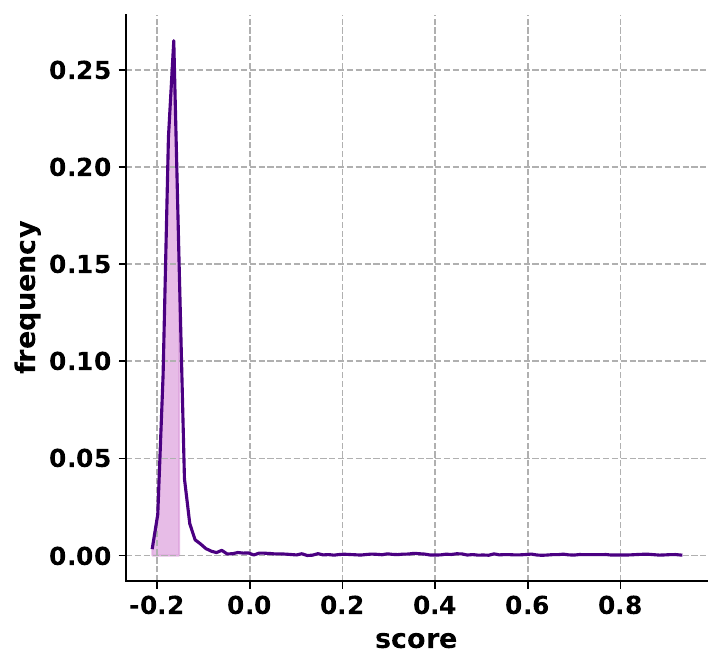}
\label{fig:epoch50}
\end{minipage}
}
\subfigure[Overall score]{
\begin{minipage}[t]{0.23\linewidth}
\centering
\includegraphics[width=\linewidth]{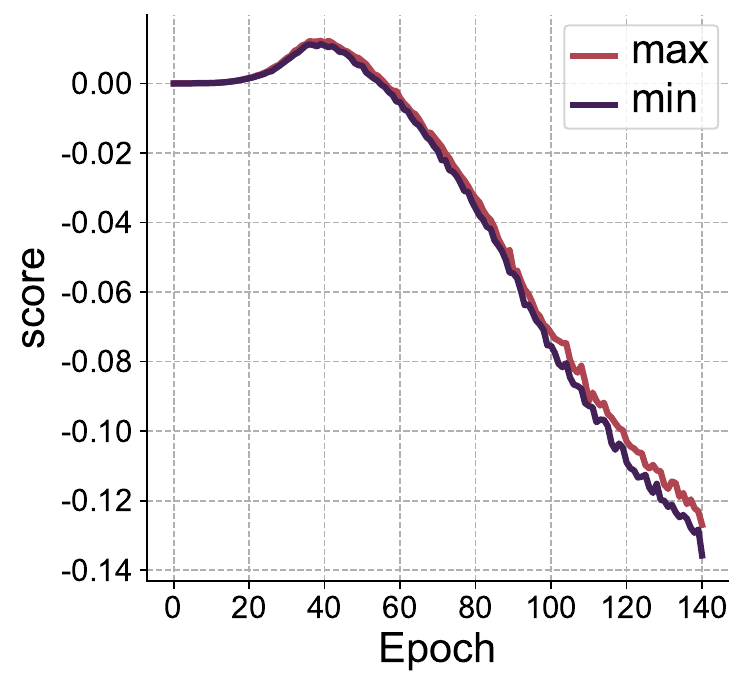}
\label{fig:overall}
\end{minipage}
}
\caption{Experimental results to demonstrate the phenomenon of ambiguous trap in DNS.}
\label{fig:trap}
\end{figure*}

\section{Limitations of the Two-Pass Approach}
In this section, we first formulate the problem of implicit CF and then explain ambiguous trap and information discrimination via intuitive experiments. 
We consider the Last.fm and Amazon-Baby datasets in experiments. A comprehensive description of the data is provided in Section~\ref{database}.
\subsection{Implicit CF}
We denote the set of historical interactions by $\mathcal{O}^{+} = \{(u, i^+) \mid u \in \mathcal{U}, i^+ \in \mathcal{I}\}$, where $\mathcal{U}$ and $\mathcal{I}$ are the set of users and the set of items, respectively. The most common implicit CF paradigm is to learn user and item representations ($\mathbf{e}_u$ and $\mathbf{e}_i$) from the historical interactions and then predict the scores of unobserved items to recommend the top-K items. The BPR loss function is widely used to optimize the model:
\begin{equation}
\mathcal{L}_{\mathrm{BPR}} = \sum_{\substack{(u, i^+) \in \mathcal{O}^+ \\ }} -\ln [\sigma(s(\mathbf{e}_u ,\mathbf{e}_{i^+}) - s(\mathbf{e}_u ,\mathbf{e}_{i^-}))],
\end{equation}
where $\sigma(\cdot)$ is the sigmoid function, and $s(\cdot,\cdot)$ is a score function (e.g., the inner product) that measures the similarity between the user and item representations.
Here $i^-$ is a negative sample selected by a sampling strategy. \emph{Our goal is to design a negative sampling strategy that is generic to different CF models.} 
Following previous studies~\cite{CLJ22,RFG12,WYZ17}, without loss of generality, we consider matrix factorization with Bayesian personalized ranking (MF-BPR)~\cite{RFG12} as the basic CF model to illustrate ANS.

\subsection{Ambiguous Trap}
We choose DNS~\cite{ZCW13}, which is the most representative hard negative sampling method, to train an MF-BPR model on the Last.fm dataset and calculate the scores of unobserved user-item pairs in different training periods. Figure~\ref{fig:epoch0}, \ref{fig:epoch30}, \ref{fig:epoch50} demonstrates the frequency distributions of the scores at different epochs. The lowest 80\% of the scores are emphasized by the pink shade. We can observe that as training progresses, more and more scores are concentrated in the low-score region, meaning that the vast majority of unobserved items are easy negative samples. Recall that the first pass samples a fixed number of negative items by a uniform distribution. Randomly sampling a small number of negative items in the first pass is difficult to include useful hard negative samples, which, in turn, substantially limits the efficacy of the second pass. We call this phenomenon \textit{ambiguous trap}.

To further demonstrate the existence of ambiguous trap, in Figure~\ref{fig:overall}, we plot the min-max normalizing maximum and minimum scores of the sampled negative items in the first pass on the Last.fm dataset. 
It can be seen that the difference between the maximum score and the minimum score is consistently small, suggesting that randomly sampling a small number of negative items makes the hardness of the negative samples obtained from DNS far from ideal. 
Note that a straightforward attempt to mitigate ambiguous trap is to substantially increase the sample size in the first pass. 
However, it is inevitably at the cost of substantial time and space overhead~\cite{CLJ22}. Inspired by contrastive learning~\cite{YYX22,WWF21,KSP20}, we propose to augment the sampled negative items to increase their hardness. 

\subsection{Information Discrimination}
\label{infor}
\begin{figure}[t]
\center
\subfigure[Last.fm]{
\begin{minipage}[t]{0.4\linewidth} 
\centering
\includegraphics[width=\linewidth]{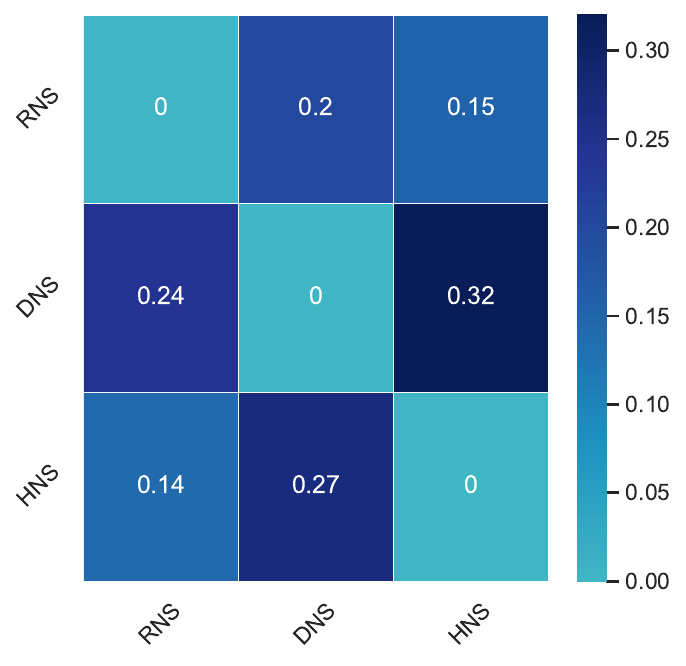}
\label{fig:per_lastfm}
\end{minipage}
}
\subfigure[Amazon-Beauty]{
\begin{minipage}[t]{0.4\linewidth}
\centering
\includegraphics[width=\linewidth]{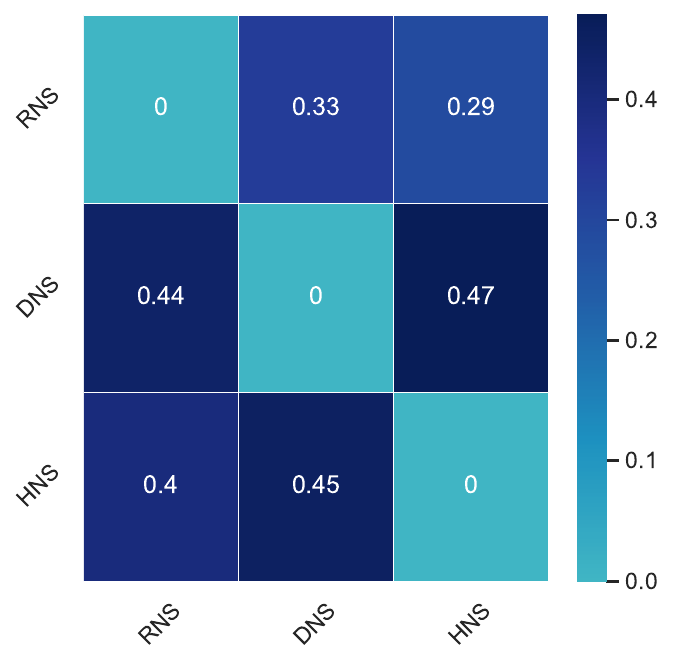}
\label{fig:per_beauty}
\end{minipage}
}
\caption{PER values of different pairs of negative sampling methods.}
\label{fig:per}
\end{figure}

In the second pass, most existing studies overly focus on high-score negative items and largely neglect low-score negative items, which also contain critical, unique information to improve model performance. Overemphasizing high-score items as negative samples may result in worse model performance.
Several studies~\cite{CLJ22,SCF23} have made efforts to assign lower sampling probabilities to low-score items using algorithms like softmax and its derivatives. However, selecting those  significantly lower-score items in comparison to others, remains a challenge. We call this behavior \textit{information discrimination}.
To verify the existence of information discrimination, we introduce a new measure named Pairwise Exclusive-hit Ratio (PER)~\cite{KLK22} to CF, which is used to compare the difference between the information learned by two different methods. More specifically, $\operatorname{PER}(x,y)$ quantifies the information captured by the method $x$ but not by $y$ via
\begin{equation}
\operatorname{PER}(x , y)=\frac{\left|\mathcal{H}_{x}-\mathcal{H}_{y}\right|}{\left|\mathcal{H}_{x}\right|},
\end{equation}
\noindent where $\mathcal{H}_x$ denotes the set of test interactions correctly predicted by the method $x$.
Next, we choose two representative negative sampling methods, RNS~\cite{DQH19} and DNS~\cite{ZCW13} to train MF-BPR models and calculate the PER between them. Recall that DNS is more likely to select high-score negative items while RNS uniformly randomly selects negative items irrespective of their scores. 
The obtained results are depicted in Figure~\ref{fig:per} (excluding HNS data for the current analysis). We can observe that: (1) the DNS strategy can indeed learn more information, confirming the benefits of leveraging hard negative samples to form a tighter decision boundary. 
(2) The values of $\operatorname{PER}(\textrm{RNS}, \textrm{DNS})$ on two datasets (0.2 and 0.33) indicate that even the simple RNS strategy can still learn rich information that is not learned by DNS. In other words, the easy negative items overlooked by DNS are still valuable for CF. Such an information discrimination problem inspires us to understand a negative item's hardness from a more fine-granular perspective in order to extract more useful information.

\section{Methodology}
\begin{figure*}[t]
  \centering
  \includegraphics[width=0.8\linewidth]{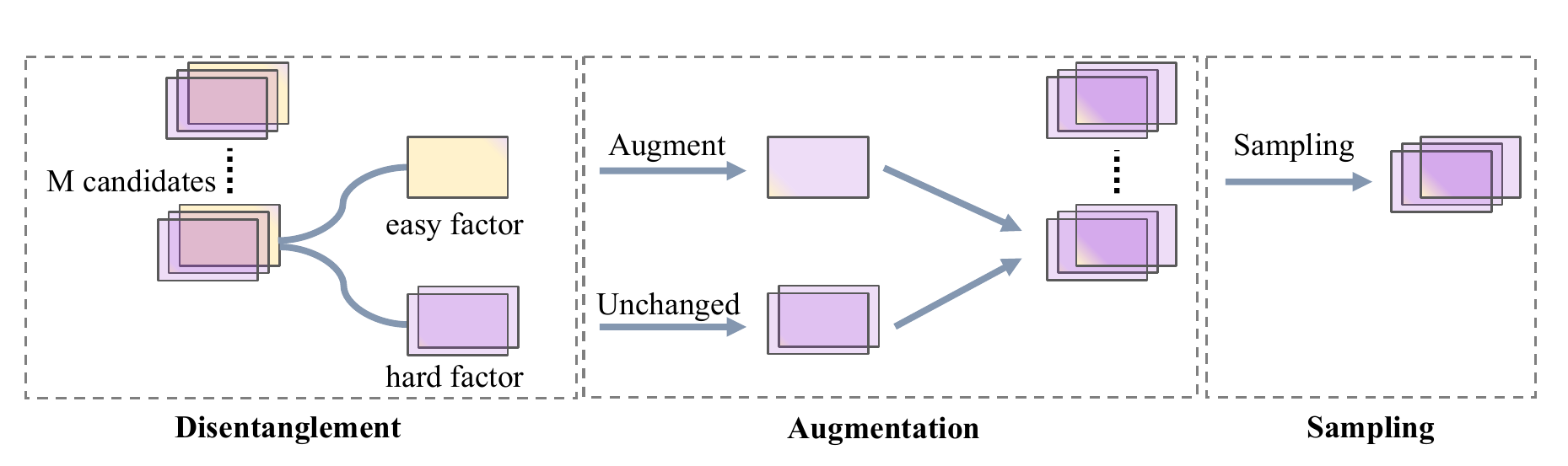}
  \caption{The workflow of augmented negative sampling (ANS) paradigm.}
  \label{fig:model}
\end{figure*}

Driven by the aforementioned limitations, we propose a novel generic augmented negative sampling (ANS) paradigm, which consists of three major steps: \emph{disentanglement}, \emph{augmentation}, and \emph{sampling}. The disentanglement step learns an item's hard and easy factors; the augmentation step adds regulated noise to the easy factor so as to increase the item's hardness; the sampling strategy selects the final negative samples based on a new metric we propose. The workflow of ANS is illustrated in Figure~\ref{fig:model}. Note that these steps can be implemented by different methods and thus the overall paradigm is generic. We present a possible instantiation in the following sections. 

\subsection{Disentanglement}
\label{disen}
To understand a negative item's hardness from a more fine-granular perspective, we propose to disentangle its embedding into hard and easy factors (i.e., a set of dimensions of the embedding vector), where hardness is defined by whether a negative item has similar values to the corresponding user in the given factor. Similarly, we follow the two-pass approach to first randomly sample $M$ ($M \ll \left\vert \mathcal{I} \right \vert$) items from the unobserved items to form a candidate negative set $\mathcal{E}$. We design a gating module to identify which dimensions of a negative item $\mathbf{e}_{n} \in \mathbb{R}^{d}$ in $\mathcal{E}$ are hard with respect to user $\mathbf{e}_{u}\in \mathbb{R}^{d}$ via
\begin{equation}
    gate_{hard} =\sigma\left(\mathbf{W}_{item} \mathbf{e}_{n} \odot \mathbf{W}_{user} \mathbf{e}_{u}\right),
\end{equation}
where $gate_{hard} \in \mathbb{R}^{d}$ gives the weights of different dimensions. The sigmoid function $\sigma(\cdot)$ maps the values to $(0, 1)$.
$\mathbf{W}_{item}\in \mathbb{R}^{d \times d}$ and $\mathbf{W}_{user}\in \mathbb{R}^{d \times d}$ are linear transformations used to ensure that the user and item embeddings are in the common latent space~\cite{MZW21}.
$\odot$ is the element-wise product, which measures the similarity between $\mathbf{e}_i$ and $\mathbf{e}_u$ in each dimension~\cite{WHW19}. 
After obtaining the weights $gate_{hard}$, we adopt the element-wise product to extract the hard factor $\mathbf{e}_n^{hard}$. The easy factor $\mathbf{e}_n^{easy}$ is then calculated via element-wise subtraction~\cite{HWN22}.
\begin{equation}
\begin{split}
\begin{aligned}
    \mathbf{e}_n^{hard} &= \mathbf{e}_n \odot gate_{hard}, \\
    \mathbf{e}_n^{easy} &= \mathbf{e}_n - \mathbf{e}_n^{hard}.
    \label{eq:easy}
\end{aligned}
\end{split}
\end{equation}

Due to the lack of ground truth for hard and easy factors, it is inherently difficult to guarantee the credibility of the disentanglement. Inspired by its superiority in unsupervised scenarios~\cite{ZGC22,KTW20}, we propose to adopt contrastive learning to guide the disentanglement.
By definition, the hard factor of a negative item should be more similar to the user, while the easy factor should be the opposite. Therefore, given a score function $s(\cdot,\cdot)$ to calculate the similarity between a pair, we design a contrastive loss $\mathcal{L}_c$ as
\begin{equation}
\begin{split}
\begin{aligned}
    \mathcal{L}_c = \sum_{{e}_n \in \mathcal{E}} s(\mathbf{e}_u, \mathbf{e}_n^{easy}) - s(\mathbf{e}_u ,  \mathbf{e}_n^{hard}).
    \label{eq:neg}
\end{aligned}
\end{split}
\end{equation}

However, optimizing only $\mathcal{L}_c$ may lead to a trivial solution: including all dimensions as the hard factor. Therefore, we introduce another loss with the auxiliary information from positive items. We adopt a similar operation with the same weights $gate_{hard}$ to obtain the corresponding positive item factor $\mathbf{e}_p^{\prime}$ and $\mathbf{e}_p^{\prime\prime}$:
\begin{equation}
\begin{split}
\begin{aligned}
    \mathbf{e}_p^{\prime} &= \mathbf{e}_p \odot gate_{hard}, \\
    \mathbf{e}_p^{\prime\prime} &= \mathbf{e}_p - \mathbf{e}_p^{\prime}.
    \label{eq:phe}
\end{aligned}
\end{split}
\end{equation}
This is particularly important as $\mathbf{e}_n^{easy}$ may emphasize the first 48 dimensions while $\mathbf{e}_n^{hard}$ may emphasize the last 14.
In order to ensure coherence in subsequent operations, it is imperative to maintain the correspondence of dimensions. For ease of understanding, reader can directly regard them as positive samples.
Intuitively, $\mathbf{e}_n^{hard}$ should be more similar to positive since it is difficult for users to discern it as a negative sample (both have a high similarity to the user). However, this signal is not entirely reliable, as it may not accurately reflect a user's level of interest. 
Therefore, instead of relying on a stringent constraint like Equation~\ref{eq:neg}, we introduce another disentanglement loss $\mathcal{L}_d$ as
\begin{equation}
\begin{split}
\begin{aligned}
    \mathcal{L}_d =\sum_{{e}_n \in \mathcal{E}} \left\|\mathbf{e}_p^{\prime}-\mathbf{e}_n^{hard}\right\|^{2} + s(\mathbf{e}_p^{\prime\prime} , \mathbf{e}_n^{easy}),
    \label{eq:lt}
\end{aligned}
\end{split}
\end{equation}
where the Euclidean distance is used to measure the similarity between $\mathbf{e}_p^{\prime}$ and $\mathbf{e}_n^{hard}$ while the score function is used to measure the similarity between $\mathbf{e}_p^{easy}$ and $\mathbf{e}_n^{\prime\prime}$. This is because we want to leverage only reliable hardness while we can be more lenient with the easy part.

\subsection{Augmentation}
\label{aug}
Next, we propose an augmentation module to create synthetic negative items which are more similar to the corresponding positive items. After the disentanglement step, we have obtained the hard and easy factors of a negative item, where the hard factor contains more useful information for model training. Therefore, our goal is to augment the easy factor to improve model performance. However, existing augmentation techniques fail to carefully regulate and quantify the augmentation needed to approximate positive items while still being negative. To this end, we propose to regulate the augmentation from two different aspects:

\textbf{Direction:} Intuitively, the direction of the augmentation on a negative item should be towards the corresponding positive item. Therefore, we first calculate the difference $\mathbf{e}_{dif}$ between the factor of the positive item $\mathbf{e}_p^{\prime\prime}$ and the easy factor of the negative item $\mathbf{e}_n^{easy}$: 
\begin{equation}
\begin{split}
\begin{aligned}
    \mathbf{e}_{dif} &= \mathbf{e}_p^{\prime\prime} - \mathbf{e}_n^{easy}.
\end{aligned}
\end{split}
\end{equation}

A first attempt is to directly make $\mathbf{e}_{dif}$ as the direction of the augmentation. 
This design is less desirable because (1) it introduces too much positive information, which may turn the augmented negative item into positive. (2) It contains too much prior information (i.e., the easy factor is identical to that of the positive item), which can lead to the model collapse problem~\cite{SVR17,TT20}. 
Inspired by~\cite{YYX22}, we carefully smooth the direction of augmentation by extracting the quadrant information $\mathbf{e}_{dir}$ of $\mathbf{e}_{dif}$ with the sign function $sgn(\cdot)$:
\begin{equation}
\begin{split}
\begin{aligned}
    \mathbf{e}_{dir} &= sgn(\mathbf{e}_{dif}).
\end{aligned}
\end{split}
\end{equation}
The direction $\mathbf{e}_{dir} \in \mathbb{R}^{d}$ effectively compresses the embedding augmentation space into a quadrant space, which provides essential direction information without having the aforementioned issues.

\textbf{Magnitude:} Magnitude determines the strength of augmentation. Several studies~\cite{HHD18} have shown that when the perturbation to the embedding is overly large, it will dramatically change its original semantics. Therefore, we need to carefully calibrate the magnitude of the augmentation. We design a two-step approach to generate a regulated magnitude $\Delta \in \mathbb{R}^{d}$. We first consider the \textit{distribution} of $\Delta$. As $\Delta$ is a noise embedding, we adopt a uniform distribution on the interval $[0, 0.1]$. The uniform distribution introduces a certain amount of randomness, which is beneficial to improve the robustness of the model.
Second, we restrict $\Delta$ to be smaller than a \textit{margin}. Instead of using a static scalar~\cite{MZW21}, we dynamically set the margin by calculating the similarity between the hard factors of the negative and corresponding positive items. The intuition is that a higher similarity between the hard factors suggests that the negative item already contains much useful information, and thus we should augment it with a smaller magnitude. Finally, the regulated magnitude $\Delta$ is calculated via
\begin{equation}
\begin{split}
\begin{aligned}
 \|\Delta\| &\le \sigma\left(\frac{1} {\mathbf{W}\left(\mathbf{e}_{n}^{hard} \odot \mathbf{e}_{p}^{\prime}\right)}\right),
\end{aligned}
\end{split}
\end{equation}
where the element-wise product $\odot$ is used to calculate the similarity between $\mathbf{e}_{n}^{hard}$ and $\mathbf{e}_{p}^{\prime}$. The transformation matrix $\mathbf{W}\in \mathbb{R}^{1 \times d}$ is used to map the similarity vector to a scalar. The sigmoid function $\sigma(\cdot)$ helps remove the sign information to avoid interfering with the learned direction $\mathbf{e}_{dir}$.
After carefully determining the direction and magnitude, we can generate the augmented version $\mathbf{e}_{n}^{aug}$ of the negative item $\mathbf{e}_{n}$ as follows:
\begin{equation}
\mathbf{e}_{n}^{aug}=\left(\mathbf{e}_{n}^{easy}+\Delta * \mathbf{e}_{dir}\right)+\mathbf{e}_{n}^{hard}.
\end{equation}
With our design, the augmented negative item becomes more similar to the corresponding positive item without causing huge changes of the semantics, and can remain negative. We apply the above operation to every item in the candidate negative set $\mathcal{E}$ and obtain the augmented negative set $\mathcal{E}^{aug}$.

\subsection{Sampling}
After obtaining the augmented candidate negative set $\mathcal{E}^{aug}$, we need to devise a sampling strategy to select the best negative item from $\mathcal{E}^{aug}$ to facilitate model training. Existing hard negative sampling methods select the negative item with the highest score, where the score is calculated by the score function $s(\cdot,\cdot)$ with the user and negative item embeddings as input. However, as explained before, negative items with relatively low scores will be seldom selected, which leads to the information discrimination issue. Although the augmentation can already alleviate information discrimination to a certain extent, we further design a more flexible and effective sampling strategy by introducing a new metric named \textit{augmentation gain}. Augmentation gain measures the score difference before and after the augmentation. Formally, the score $score$ and the augmentation gain $score^{aug}$ are calculated as:
\begin{equation}
\begin{split}
\begin{aligned}
score &= s(\mathbf{e}_u, \mathbf{e}_n^{aug}), \\ 
score^{aug} &= s(\mathbf{e}_u, \mathbf{e}_n^{aug}) - s(\mathbf{e}_u, \mathbf{e}_n).
\end{aligned}
\end{split}
\end{equation}
The sampling module we propose considers \emph{both} $score$ and $score^{aug}$ to select the suitable negative item $\mathbf{e}_{n}^{final}$ from the augmented candidate negative set $\mathcal{E}^{aug}$. To explicitly balance the contributions between $score$ and $score^{aug}$, we introduce a trade-off parameter $\epsilon$. The final augmented negative item $\mathbf{e}_{n}^{final}$ is chosen via
\begin{equation}
\begin{split}
\begin{aligned}
\mathbf{e}_{n}^{final}=\underset{\mathbf{e}_n^{aug} \in \mathcal{E}^{aug}}{\arg \max } \left(score + \epsilon score^{aug}\right).
\end{aligned}
\end{split}
\end{equation}

\subsection{Discussion}
It is worth noting that most existing negative sampling methods can be considered as a special case of ANS. For example, DNS~\cite{ZCW13} can be obtained by removing the disentanglement and augmentation steps. MixGCF~\cite{HDD21} can be obtained by removing the disentanglement step and replacing the regulated augmentation with unconstrained augmentation based on graph structure information and positive item information. SRNS~\cite{DQY20} can be obtained by removing the augmentation step and considering variance in the sampling strategy step.

\subsection{Model Optimization}
Finally, we adopt the proposed ANS method as the negative sampling strategy and take into consideration also the recommendation loss $\mathcal{L}$ to optimize the parameters $\Theta$ of an implicit CF model (e.g., MF-BPR).
\begin{equation}
\mathcal{L} = \sum_{\substack{(u, i^+) \in \mathcal{O}^+ \\ }} -\ln [\sigma(s(\mathbf{e}_u ,\mathbf{e}_{p}) - s(\mathbf{e}_u ,\mathbf{e}_{n}^{final}))] + \gamma(\mathcal{L}_c + \mathcal{L}_d) +\lambda\Vert \Theta \Vert_2^2,
\end{equation}
\noindent where 
$\lambda$ is a hyper-parameter controlling the strength of $L_2$ regularization, and $\gamma$ is another hyper-parameter used to adjust the impact of the contrastive loss and the disentanglement loss.
Observably, our negative sampling strategy paradigm can be seamlessly incorporated into mainstream models without the need for substantial modifications.

\section{EXPERIMENT}
\label{database}
In this section, we conduct comprehensive experiments to answer the following key research questions:
\begin{itemize}[leftmargin=*]
    \item \textbf{RQ1:} How does ANS perform compared to the baselines and integrating ANS into different mainstream CF models perform compared with the original ones?
    \item \textbf{RQ2:} How accurate is the disentanglement step in the absence of ground truth?
    \item \textbf{RQ3:} Can ANS alleviate ambiguous trap and information discrimination?
    \item \textbf{RQ4:} How do different steps affect ANS’s performance?
    \item \textbf{RQ5:} How do different hyper-parameter settings (i.e., $\gamma$, $\epsilon$, and $M$) affect ANS’s performance?
    \item \textbf{RQ6:} How does ANS perform in efficiency?

\end{itemize}

\subsection{Experimental Setup}
\subsubsection{Datasets}
We consider five public benchmark datasets in the experiments: Amazon-Baby, Amazon-Beauty, Yelp2018, Gowalla, and Last.fm.  
In order to comprehensively showcase the efficacy of the proposed methodology, we have partitioned the dataset into two distinct categories for processing. For the datasets Amazon-baby, Amazon-Beauty, and Last.fm, the training set is constructed by including only the interactions that occurred on or before a specified timestamp, similar to the approach used in the state-of-the-art method DENS~\cite{LCZ23}. After reserving the remaining interactions for the test set, a validation set is created by randomly sampling 10\% of the interactions from the training set. The adoption of this strategy, as suggested by previous works~\cite{WDH20,CLJ22,MZW21}, provides the benefit of preventing data leakage. 
For Yelp and Gowalla, we have followed the conventional practice of utilizing an 80\% training set, 10\%  test set, and 10\%  validation set. These datasets have different statistical properties, which can reliably validate the performance of a model~\cite{CCC22}. Table~\ref{tab:statistics} summarizes the statistics of the datasets. 
\begin{table}[t]
\begin{center}
\caption{The statistics of the datasets used in the experiments.}
\vspace{-3mm}
\begin{tabular}{@{}lccccc@{}}
\toprule
\textbf{Dataset} & \textbf{Users} & \textbf{Items} & \textbf{Interactions} & \textbf{\makecell[c]{Cutting \\ Timestamp}} \\ \midrule
Amazon-Baby & 4,314 & 5,854 & 42,657 & Apr. 1, 2014 \\ 
Amazon-Beauty & 7,670 & 10,887 & 82,297  & Apr. 1, 2014\\ 
Last.fm & 640 & 4,165 & 120,975 & Apr. 1, 2010\\
Yelp & 31,669 & 38,049 & 1,561,406 & -  \\ 
Gowalla & 29,859 & 40,982 & 1,027,370 & -  \\
\bottomrule
\end{tabular}
\label{tab:statistics}
\end{center}
\vspace{-5mm}
\end{table}

\subsubsection{Baseline Algorithms}
To demonstrate the effectiveness of the proposed ANS method, we compare it with several representative state-of-the-art negative sampling methods.
\begin{itemize}[leftmargin=*]
    \item \textbf{RNS}~\cite{RFG12}: Random negative sampling (RNS) adopts a uniform distribution to sample unobserved items. 
    \item \textbf{DNS}~\cite{ZCW13}: Dynamic negative sampling (DNS) adaptively selects items with the highest score as the negative samples.
    \item \textbf{SRNS}~\cite{DQY20}: SRNS introduces variance to avoid the false negative item problem based on DNS.
    \item \textbf{MixGCF}~\cite{HDD21}: MixGCF injects information from positive and graph to synthesizes harder negative samples.
    \item \textbf{DENS}~\cite{LCZ23}: DENS disentangles relevant and irrelevant factors of items and designs a factor-aware sampling strategy.

\end{itemize}

To further validate the effectiveness of our proposed methodology, we have integrated it with a diverse set of representative models.
\begin{itemize}[leftmargin=*]
    \item \textbf{NGCF}~\cite{WHW19}: NGCF employs a message-passing scheme to effectively leverage the high-order information.
    \item \textbf{LightGCN}~\cite{HDW20}: LightGCN adopts a simplified approach by eliminating the non-linear transformation and instead utilizing a sum-based pooling module to enhance its performance.
    \item \textbf{SGL}~\cite{WWF21}: SGL incorporates contrastive learning. The objective is to enhance the agreement between various views of the same node, while minimizing the agreement between views of different nodes.
\end{itemize}

\subsubsection{Implementation Details}
Similar to previous studies~\cite{CLJ22,RFG12,WYZ17}, we consider MF-BPR~\cite{RFG12} as the basic CF model. For a fair comparison, the size of embeddings is fixed to 64, and the embeddings are initialized with Xavier~\cite{GB10} for all methods. We use Adam~\cite{KB14} to optimize the parameters with a default learning rate of 0.001 and a default mini-batch size of 2048. The $L_2$ regularization coefficient $\lambda$ is set to $10^{-4}$. The size of the candidate negative item set $M$ is tested in the range of $\{2, 4, 8, 16, 32\}$. The weight $\gamma$ of the contrastive and disentanglement losses and $\epsilon$ of the augmentation gain are both searched in the range of $[0, 1]$. In order to guarantee the replicability, our approach is implemented by the RecBole v1.1.1 framework~\cite{ZMH21}. We conducted statistical tests to evaluate the significance of our experimental results.

\begin{table*}[t]
\begin{center}
\caption{The performance of different negative sampling strategies on MF-BPR. The best results are boldfaced, and the second-best results are underlined.}

\begin{tabular}{@{}lcccccccccccc@{}}
\toprule
\multirow{2}{*}{\textbf{Dataset}} & \multirow{2}{*}{\textbf{Method}} &\multicolumn{3}{c}{Top-10} &\multicolumn{3}{c}{Top-15} &\multicolumn{3}{c}{Top-20}\\ 
\cmidrule(lr){3-5} \cmidrule(lr){6-8} \cmidrule(lr){9-11} & & Hit Ratio & Recall & NDCG & Hit Ratio & Recall & NDCG & Hit Ratio & Recall & NDCG \\ \midrule
\multirow{4}{*}{\textbf{Baby}} & RNS & 0.0353 & 0.0126 & {0.0086} & 0.0455 & 0.0176 & 0.0098 & 0.0603 & 0.0232 & 0.0116 \\
& DNS  & {0.0366} & {0.0133} & {0.0086} & {0.0507} & {0.0189} & {0.0103} & 0.0625 & {0.0238} & 0.0118 \\ 
& SRNS & 0.0361 & 0.0131 & 0.0085 & 0.0487 & 0.0183 & 0.0102 & {0.0630} & {0.0235} & {0.0119} \\
& MixGCF & {0.0363} & {0.0126} & {0.0083} & {0.0501} & {0.0178} & {0.0100} & {0.0656} & {0.0248} & {0.0121} \\ 
& DENS & \underline{0.0514} & \underline{0.0182} & \underline{0.0113}& \underline{0.0687} & \underline{0.0260} & \underline{0.0140} & \underline{0.0843} & \underline{0.0320} & \underline{0.0158} \\
& ANS  & \textbf{0.0585} & \textbf{0.0214}& \textbf{0.0125} & \textbf{0.0766} & \textbf{0.0288} & \textbf{0.0150} & \textbf{{0.0961}} & \textbf{{0.0352}} & \textbf{{0.0171}} \\

\midrule
\multirow{4}{*}{\textbf{Beauty}} & RNS & 0.0354 & 0.0151 & 0.0099 & 0.0449 & 0.0201 & 0.0114 & 0.0509 & 0.0229 & 0.0122 \\
& DNS & {0.0396} & {0.0171} & {0.0114} & {0.0501} & 0.0218 & {0.0130} & {0.0596} & {0.0262} & {0.0143} \\
& SRNS & 0.0395 & 0.0168 & {0.0114} & \underline{0.0504} &{0.0219} & 0.0129 & 0.0584 & 0.0260 & 0.0142 \\ 
& MixGCF & {0.0363} & {0.0126} & {0.0083} & {0.0501} & {0.0178} & {0.0100} & \underline{0.0656} & {0.0248} & {0.0121} \\ 
& DENS & \underline{0.0397} & \underline{0.0171} & \underline{0.0116} &  0.0501 & \underline{0.0223} & \underline{0.0132} & 0.0597 & \underline{0.0266} & \underline{0.0146} \\
& ANS & \textbf{0.0501}& \textbf{0.0210} & \textbf{{0.0151}} & \textbf{0.0650} & \textbf{{0.0276}} & \textbf{0.0173} & \textbf{0.0747} & \textbf{0.0323} &\textbf{0.0188} \\

\midrule
\multirow{4}{*}{\textbf{Last.fm}} & RNS & 0.2800 & 0.0656 & 0.0914 & 0.3113 & 0.0810 & 0.0908 & 0.3304 & 0.0929 & 0.0909 \\

& DNS & {0.2957} & {0.0731} & 0.0993 & \textbf{0.3409} & {0.0902} & {0.0990} & \underline{0.3652} & {0.1067} & {0.0999} \\
& SRNS & 0.2939 & \textbf{{0.0980}} & \underline{0.1008} & \underline{0.3374} & 0.0901 & 0.0975 & 0.3617 & 0.1052 & 0.0988 \\
& MixGCF & {0.2699} & 0.706 & {0.0872} & {0.2965} & {0.0891} & {0.0877} & {0.3251} & {0.1101} & {0.0903} \\
& DENS & \underline{0.3009} & 0.0711 & {0.0997} & {0.3426} & \underline{0.0933} & \underline{0.0992} & \textbf{{0.3843}} & \underline{0.1105} & \underline{0.1015} \\
& ANS & \textbf{0.3128} &  \underline{0.0902} & \textbf{0.1031} & {0.3333} & \textbf{0.1067} & \textbf{0.1013} & {0.3619} & \textbf{0.1282} & \textbf{0.1045} \\

\midrule
\multirow{4}{*}{\textbf{Yelp}} & RNS & 0.1537 & 0.0427 & 0.0334 & 0.2005 & 0.0574 & 0.0385 & 0.2398 & 0.0717 & 0.0432 \\
& DNS & 0.1843 & 0.0514 & 0.0409 & 0.2359 & 0.0694 & 0.0471 & 0.2773 & 0.0848 & 0.0521\\
& SRNS & 0.1869 & 0.0530 & 0.0420 & 0.2395 & 0.0711 & 0.0482 & 0.2810 & 0.0869 & 0.0535 \\
& MixGCF & \underline{0.1891} & \underline{0.0538} & 0.0424 & \underline{0.2412} & 0.0716 & 0.0485 & \underline{0.2830} & 0.0871 & 0.0536 \\ 
& DENS& 0.1876 & \underline{0.0538} & \underline{0.0425} & 0.2391 & \underline{0.0717} & \underline{0.0487} & 0.2817 & \underline{0.0877} & \underline{0.0539} \\
& ANS & \textbf{0.1940} & \textbf{0.0552} & \textbf{0.0439} & \textbf{0.2458} & \textbf{0.0732} & \textbf{0.0500} & \textbf{0.2894} & \textbf{0.0902} & \textbf{0.0556} \\
\midrule
\multirow{4}{*}{\textbf{Gowalla}} & RNS & 0.1989 & 0.0938 & 0.0674 & 0.2420 & 0.1182 & 0.0746 & 0.2763 & 0.1389 & 0.0804 \\
& DNS & 0.2380 & 0.1166 & 0.0834 & 0.2878 & 0.1472 & 0.0923 & 0.3238 & 0.1708 & 0.0989\\
& SRNS & \textbf{0.2462} & 0.1201 & 0.0867 & \underline{0.2939} & \underline{0.1499} & \underline{0.0954} & 0.3319 & 0.1746 & 0.1023 \\
& MixGCF & 0.2401 & 0.1189 & 0.0853 & 0.2881 & 0.1484 & 0.0940 & 0.3254 & 0.1728 & 0.1008 \\ 
& DENS& \underline{0.2457} & \underline{0.1203} & \underline{0.0867} &  0.2929 & 0.1495 & 0.0953 & \underline{0.3322} & \underline{0.1765} & \underline{0.1027} \\
& ANS & \textbf{0.2462} & \textbf{0.1210} & \textbf{0.0874} & \textbf{0.2972} & \textbf{0.1524} & \textbf{0.0966} & \textbf{0.3350} & \textbf{0.1778} & \textbf{0.1036} \\

\bottomrule
\end{tabular}
\label{tab:mf}
\end{center}
\end{table*}

\begin{table*}[t]
\caption{Experimental results of different CF model with (w) or without (w/o) ANS. The best results are boldfaced.}
\vspace{-2mm}
\label{tab:with}
\begin{tabular}{@{}clllllllllllll@{}}
\toprule
\multirow{2}{*}{\textbf{Datasets}} & \multirow{2}{*}{\textbf{Metric}} & \multicolumn{2}{c}{\textbf{MF-BPR}} & \multicolumn{2}{c}{\textbf{LightGCN}} & \multicolumn{2}{c}{\textbf{NGCF}} & \multicolumn{2}{c}{\textbf{SGL}} &  \\ \cmidrule(l){3-10} 
 &  & w/o & w & w/o & w & w/o & w & w/o & w  \\ \midrule 
\multirow{6}{*}{Gowalla} 
 & Recall@10 & 0.0938 & \textbf{0.1210} & 0.1280 & \textbf{0.1360} & 0.1011 & \textbf{0.1115} & 0.1331 & \textbf{0.1347}  \\
 & Recall@20 & 0.1389 & \textbf{0.1778} & 0.1849 & \textbf{0.1967} & 0.1490 & \textbf{0.1642} & 0.1897 & \textbf{0.1961}  \\
 & Hit@10 & 0.1989 & \textbf{0.2462} & 0.2553 & \textbf{0.2712} & 0.2110 & \textbf{0.2286} & 0.2667 & \textbf{0.2688}  \\
 & Hit@20 & 0.2763 & \textbf{0.3350} & 0.3432 & \textbf{0.3626} & 0.2900 & \textbf{0.3130} & 0.3519 & \textbf{0.3637}  \\
 & NDCG@10 & 0.0674 & \textbf{0.0874} & 0.0917 & \textbf{0.0981} & 0.0722 & \textbf{0.0795} & 0.0966 & \textbf{0.0977} \\
 & NDCG@20 & 0.0804 & \textbf{0.1036} & 0.1079 & \textbf{0.1155} & 0.0859 & \textbf{0.0945} & 0.1128 & \textbf{0.1154}  \\

\midrule 

\multirow{6}{*}{Yelp} 
 & Recall@10 & 0.0427 & \textbf{0.0537} & 0.0543 & \textbf{0.0639} & 0.0439 & \textbf{0.0477} & 0.0612 & \textbf{0.0622}  \\
 & Recall@20 & 0.0717 & \textbf{0.0875} & 0.0884 & \textbf{0.1044} & 0.0738 & \textbf{0.0787} & 0.0990 & \textbf{0.1014}  \\
 & Hit@10 & 0.1537 & \textbf{0.1887} & 0.1896 & \textbf{0.2183} & 0.1593 & \textbf{0.1679} & 0.2101 & \textbf{0.2117}  \\
 & Hit@20 & 0.2398 & \textbf{0.2822} & 0.2847 & \textbf{0.3240} & 0.2460 & \textbf{0.2550} & 0.3098 & \textbf{0.3153}  \\
 & NDCG@10 & 0.0334 & \textbf{0.0430} & 0.0432 & \textbf{0.0517} & 0.0345 & \textbf{0.0377} & 0.0487 & \textbf{0.0490} \\
 & NDCG@20 & 0.0432 & \textbf{0.0543} & 0.0546 & \textbf{0.0653} & 0.0447 & \textbf{0.0482} & 0.0614 & \textbf{0.0623}  \\
  \midrule 
\vspace{-5mm}
\end{tabular}
\end{table*}
\subsection{RQ1: Overall Performance Comparison}
Table~\ref{tab:mf} shows the results of training MF-BPR with different negative sampling methods. Additionally, the performance of ANS under different models is presented in Table~\ref{tab:with}. Due to space limitations, we are unable to present the results of other models using various negative sampling strategies. Nonetheless, it is worth noting that the experimental results obtained were similar with Table~\ref{tab:mf}.
We can make the following key observations:
\begin{itemize}[leftmargin=*]
    \item ANS yields the best performance on almost all datasets. In particular, its highest improvements over the strongest baselines are 29.74\%, 23.76\%, and 31.06\% in terms of $Hit~Ratio@15$, $Recall@15$, and $NDCG@15$ in Beauty, respectively.
    This demonstrates that ANS is capable of generating more informative negative samples. 
    \item  The ANS's remarkable adaptability is a noteworthy feature that allows for its seamless integration into various models. The results presented in Table~\ref{tab:with} demonstrate that the incorporation of PAN into the base models leads to improvements across all datasets.
    \item The model-aware methods always outperform the model-agnostic methods (RNS). In general, model-agnostic methods are difficult to guarantee the quality of negative samples. Leveraging various information from the underlying model is indeed a promising research direction. 
    \item Despite its simplicity, DNS is a strong baseline. This fact justifies our motivation of studying the hardness from a more fine granularity.

\end{itemize}

\subsection{RQ2: Disentanglement Performance}
\label{disen_performance}
\begin{figure}[t]
\center
\subfigure[Factor distribution in $\mathbb{R}^2$]{
\begin{minipage}[t]{0.4\linewidth} 
\centering
\includegraphics[width=\linewidth]{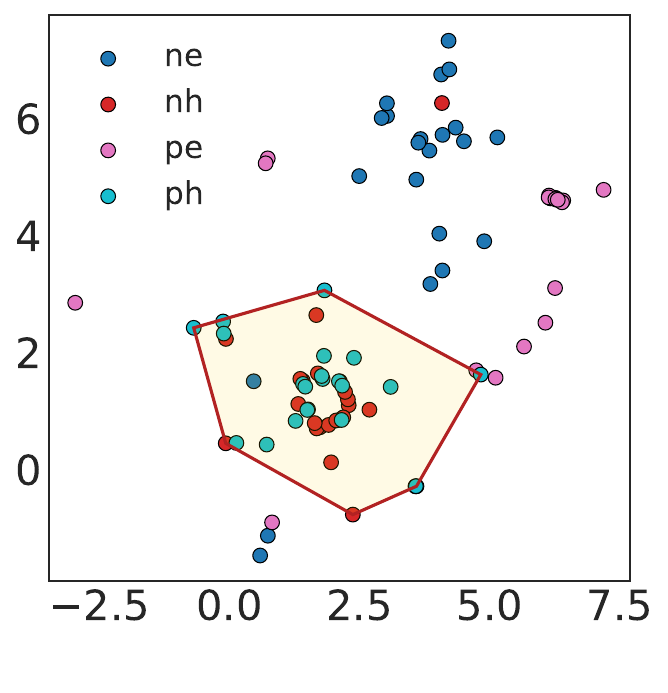}
\label{fig:bubble}
\vspace{-5mm}
\end{minipage}
}
\subfigure[Training curves]{
\begin{minipage}[t]{0.4\linewidth}
\centering
\includegraphics[width=\linewidth]{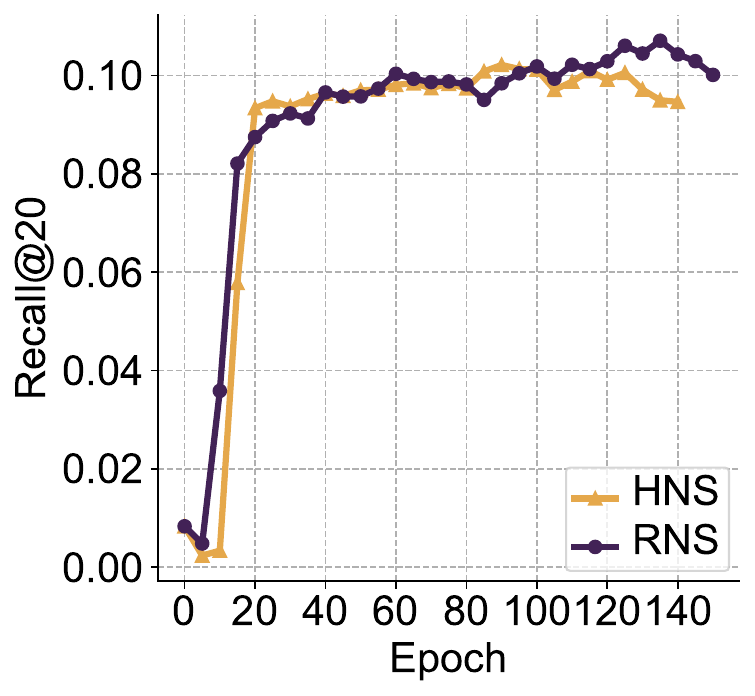}
\label{fig:hns}
\vspace{-5mm}
\end{minipage}
}
\vspace{-3mm}
\caption{(a) shows the distribution of different factors learned by ANS on Last.fm. ne, nh, pe, and ph denote the $\mathbf{e}_{n}^{easy}$, $\mathbf{e}_{n}^{hard}$, $\mathbf{e}_{p}^{\prime}$, and $\mathbf{e}_{p}^{\prime\prime}$ respectively. (b) shows the training curves on Last.fm.}
\label{fig:dis}
\vspace{-5mm}
\end{figure}
To verify the disentanglement performance, we spot-check a user and use the T-SNE algorithm~\cite{VH08} to map the disentangled factors into a two-dimensional space. The results are shown in Figure~\ref{fig:bubble}. 
We can observe that $\mathbf{e}_{n}^{hard}$ and $\mathbf{e}_{p}^{\prime}$ are clustered together, confirming that they are indeed similar. In contrast, $\mathbf{e}_{n}^{easy}$ and $\mathbf{e}_{p}^{\prime\prime}$ are more scattered, indicating that they do not carry similar information, which is consistent with our previous analysis.

The hard factors should contain most of the useful information of negative items, and thus if we use only hard factors, instead of the entire original item, to train a model, we should achieve similar performance. We validate it by two experiments. First, we plot the curves of $Recall@20$ in Figure~\ref{fig:hns}. HNS is a variant of RNS, which uses our disentanglement step to extract the hard factors of items and then uses \emph{only} hard factors to train the model. It can be observed that the performance of HNS is comparable to that of RNS. The two curves are similar, confirming that the hard factors indeed capture the most useful information of negative items. 
Second, we revisit Figure~\ref{fig:per} in Section~\ref{infor}. It can be seen that $\operatorname{PER}(\textrm{RNS}, \textrm{HNS})$ is small, which means that HNS captures most of information learned in RNS.

In summary, our disentanglement step can effectively disentangle a negative sample into hard and easy factors, which lays a solid foundation for the subsequent steps in ANS.

\subsection{RQ3: Ambiguous Trap and Information Discrimination}

\begin{figure}[t]
\center
\includegraphics[width=0.6\linewidth]{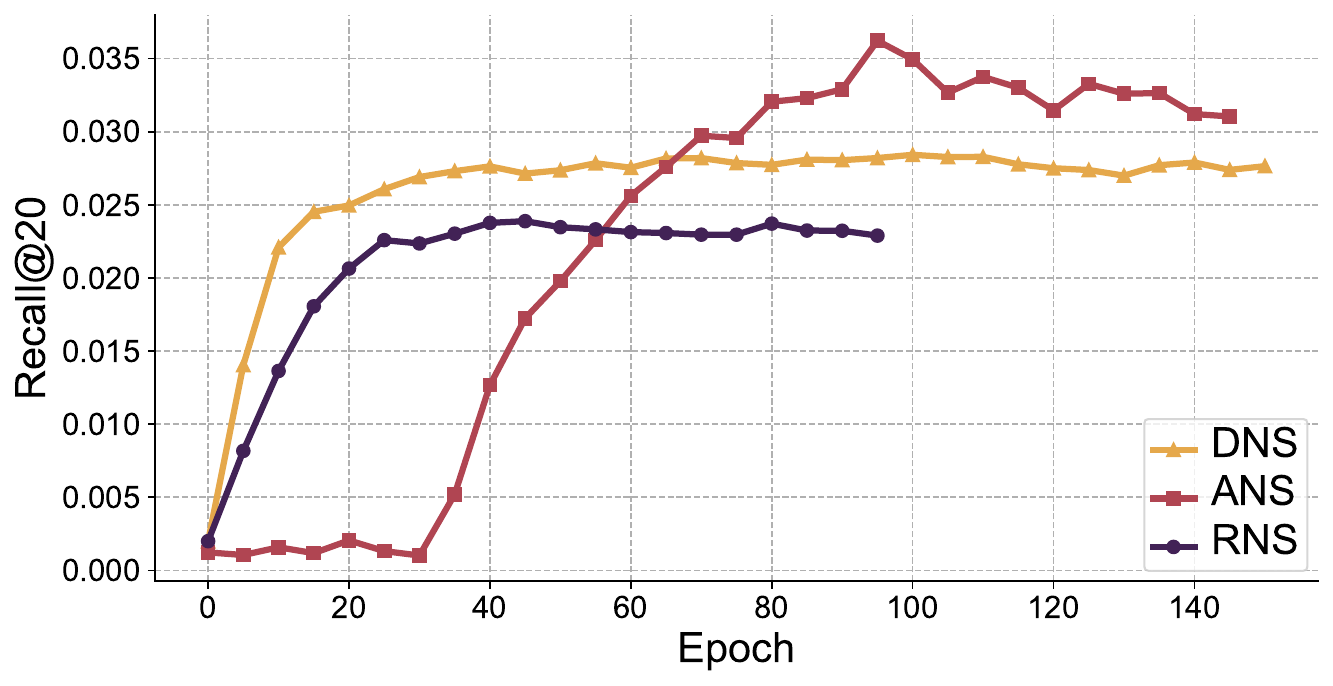}
\caption{The curves of $Recall@20$ on the baby dataset.}
\label{fig:recall}
\vspace{-3mm}
\end{figure}
\subsubsection{Ambiguous Trap}
Demonstrating how ANS can mitigate the ambiguous trap is a challenging task. A first attempt is to show the augmentation gain of the augmented negative samples. However, this idea is flawed because larger augmentation gain cannot always guarantee better performance (e.g., larger augmentation gain can be achieved by introducing a large number of false positive items). Therefore, we choose to analyze the curves of $Recall@20$ to show how ANS can mitigate the ambiguous trap, which is shown in Figure~\ref{fig:recall}. Generally, the steeper the curve is, the more information the model can learn in this epoch from negative samples.
We can observe that DNS outperforms RNS because its $Recall@20$ is always higher than that of RNS and the curve is steeper than that of RNS. It again confirms that hard negative sampling is an effective sampling strategy. In contrast, the curve of ANS exhibits distinct patterns. At the beginning (from epoch 0 to epoch 30), $Recall@20$ is low because ANS needs extra efforts to learn how to disentangle and augment negative samples. As the training process progresses (from epoch 30 to epoch 95),
ANS demonstrates a greater average gradient. This proves that ANS can generate harder synthetic negative samples, which can largely mitigate the ambiguous trap issue.

\begin{table}[t]
\begin{center}
\caption{The percentage of overlapping high-score items between DNS and ANS.}
\vspace{-2mm}
\begin{tabular}{@{}lccccc@{}}
\toprule
\textbf{Dataset} & \textbf{Baby} & \textbf{Beauty}  & \textbf{Toy} & \textbf{Last.fm}\\ \midrule
Overlapping & 47.3\% & 49.0\% & 40.9\% &  38.1\%\\ 
\bottomrule
\end{tabular}
\label{tab:frequency}
\end{center}
\vspace{-5mm}
\end{table}

\subsubsection{Information Discrimination}
As for the information discrimination problem, we have shown that the disentanglement step can effectively extract the useful information from low-score items. However, we have not shown that ANS can select more low-score negative items in the training process. To this end, we analyze the percentages of overlapping negative samples between ANS and DNS. The results are presented in Table~\ref{tab:frequency}.
Recall that DNS always chooses the negative items with the highest scores as the hard negative samples. A less than 50\% overlapping indicates that ANS does sample more low-score negative items before the augmentation and effectively alleviates the information discrimination problem.
\subsection{RQ4: Ablation Study}
We analyze the effectiveness of different components in our model, and evaluate the performance of the following variants of our model: (1) ANS without disentanglement (ANS w/o dis). (2) ANS without augmentation gain (ANS w/o gain). (3) ANS without regulated direction (ANS w/o dir). (4) ANS without regulated magnitude (ANS w/o mag). It is noteworthy to state that the complete elimination of the augmentation step is not taken into consideration due to its equivalence to DNS.

The results are presented in Table~\ref{tab:ablation}. We can observe that all components we propose can positively contribute to model performance. In particular, the results show that unconstrained augmentation (e.g., ANS w/o mag) cannot achieve meaningful performance. This fact confirms that the existing unconstrained augmentation techniques cannot be directly applied to CF.  
\begin{table}[t]
\begin{center}
\caption{Ablation Study.}
\vspace{-2mm}
\begin{tabular}{@{}lcccccc@{}}
\toprule
&\textbf{Dataset} &\multicolumn{2}{c}{\textbf{Last.fm}} &\multicolumn{2}{c}{\textbf{Beauty}} \\ \midrule
&\multirow{2}{*}{\textbf{Method}} &\multicolumn{2}{c}{Top-20} &\multicolumn{2}{c}{Top-20}\\ 
\cmidrule(lr){3-4} \cmidrule(lr){5-6}  & & Recall & NDCG & Recall & NDCG \\ \midrule
& ANS w/o dis & 0.1108 & 0.0936 & 0.0286 & 0.0155\\ 
& ANS w/o gain & 0.1269 & 0.1036 & 0.0305 & 0.0166 \\
& ANS w/o dir & 0.1025 & 0.0806 & 0.0191 & 0.0106 \\ 
& ANS w/o mag & 0.0056 & 0.0024 & 0.0018 & 0.0007 \\
& ANS & \textbf{0.1282} & \textbf{0.1045} & \textbf{0.0323} & \textbf{0.0188}\\\bottomrule
\label{tab:ablation}
\end{tabular}
\vspace{-5mm}
\end{center}
\end{table}

\subsection{RQ5: Hyper-Parameter Sensitivity}
\subsubsection{Impact of $\gamma$}
\begin{figure}[t] \centering 
\subfigure[Amazon-Baby]{\begin{minipage}[t]{0.39\linewidth}\centering\includegraphics[width=\linewidth]{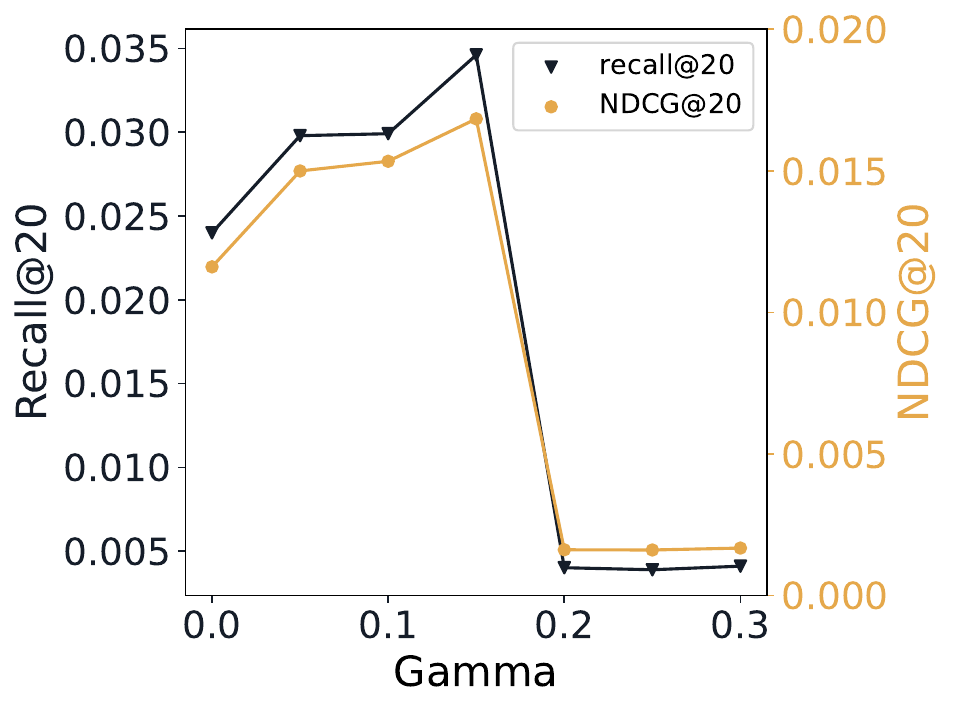}\label{fig:gamma_baby}\end{minipage}} \subfigure[Last.fm]{\begin{minipage}[t]{0.38\linewidth}\centering\includegraphics[width=\linewidth]{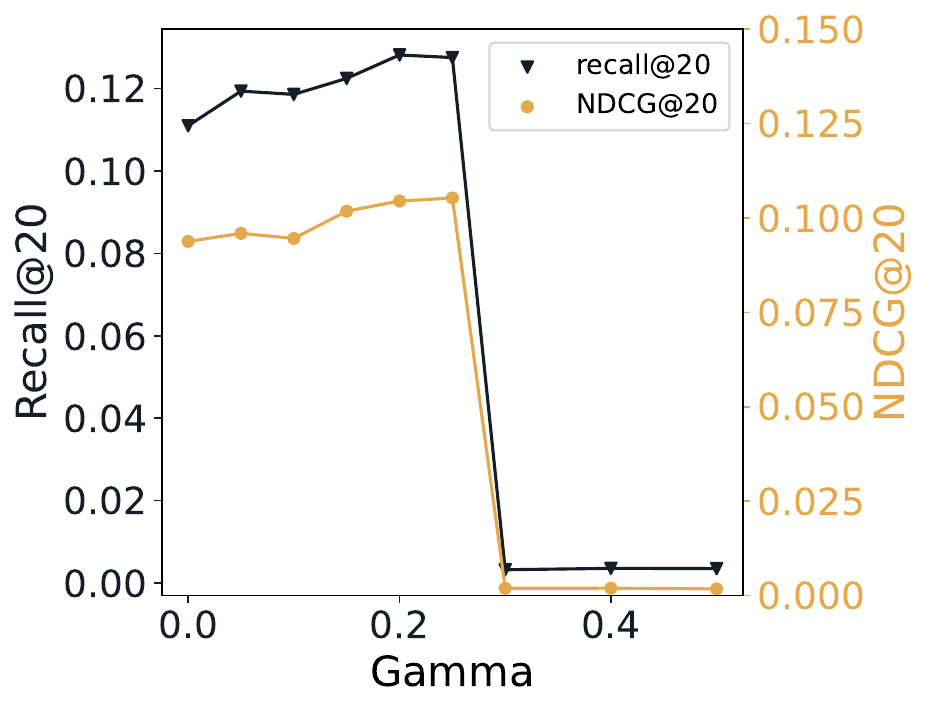}\label{fig:gamma_lastfm}\end{minipage}} 
\vspace{-5mm}
\caption{Impact of $\gamma$ in ANS.} 
\label{fig:gamma} 
\vspace{-5mm}
\end{figure}

We present the effect of the weight of the contrastive loss and disentanglement loss, $\gamma$, in Figure~\ref{fig:gamma}. As the value of $\gamma$ increases, we can first observe notable performance improvements, which proves that both loss functions are beneficial for the CF model. It is interesting to observe that once $\gamma$ becomes larger than a threshold, the performance drops sharply. This observation is expected because in this case the CF model considers the disentanglement as the primary task and ignores the recommendation task. Nevertheless, ANS can achieve reasonable performance under a relatively wide range of $\gamma$ values.

\subsubsection{Impact of $\epsilon$}
\begin{figure}[t] \centering 
\subfigure[Amazon-Baby]{\begin{minipage}[t]{0.39\linewidth}\centering\includegraphics[width=\linewidth]{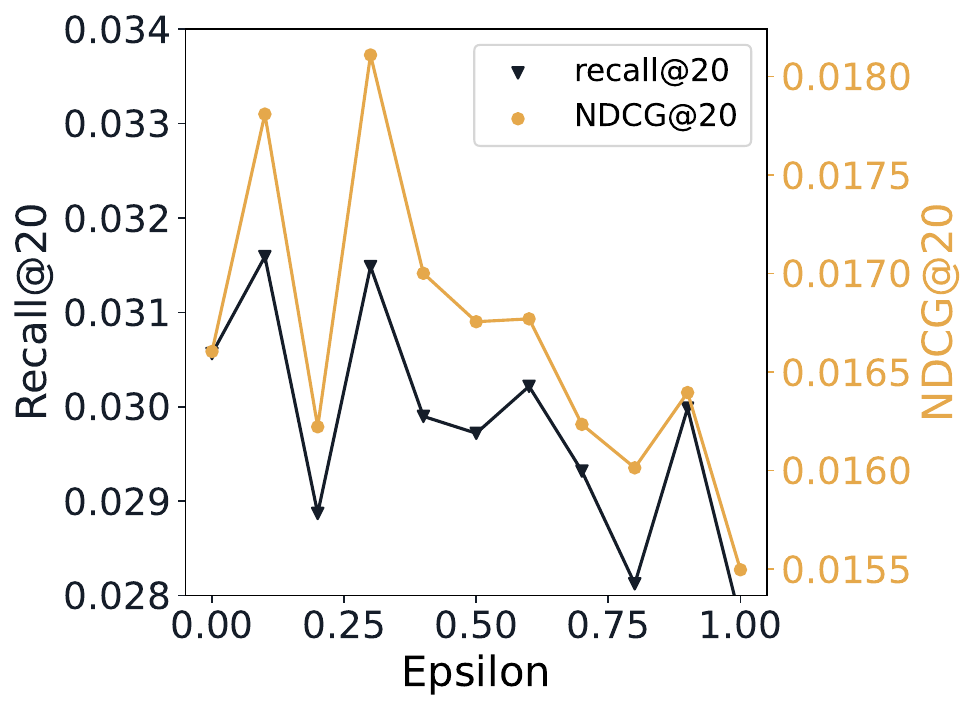}\label{fig:eps_baby}\end{minipage}} \subfigure[Last.fm]{\begin{minipage}[t]{0.38\linewidth}\centering\includegraphics[width=\linewidth]{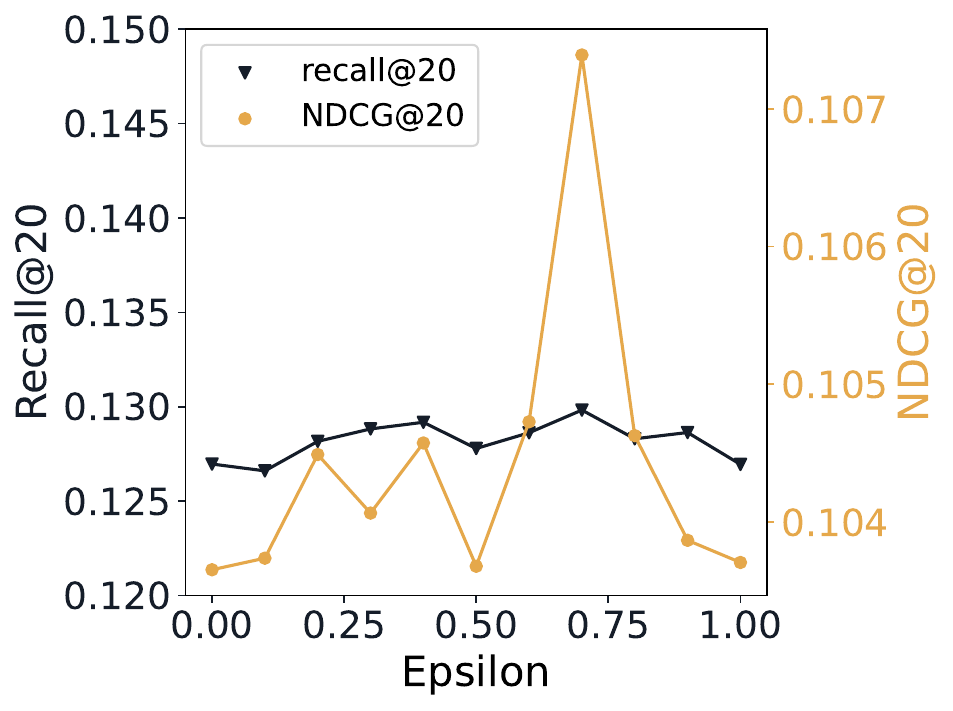}\label{fig:eps_lastfm}\end{minipage}} 
\vspace{-5mm}\caption{Impact of $\epsilon$ in ANS.} \label{fig:epsilon} 
\vspace{-5mm}
\end{figure}

Recall that $\epsilon$ is the parameter to balance the importance between the score and the augmentation gain in the sampling step. Figure~\ref{fig:eps_baby} presents the results of different $\epsilon$ values. A small $\epsilon$ value overlooks the importance of the augmentation gain and only achieves sub-optimal performance; a large $\epsilon$ value may favor an item with a lower score (but a larger score difference), which will reduce the gradient and hurt model performance. But still, ANS can obtain good performance under a relatively wide range of $\epsilon$ values. 

\subsubsection{Impact of $M$}
\begin{figure}[t] \centering 
\subfigure[Yelp]{\begin{minipage}[t]{0.38\linewidth}\centering\includegraphics[width=\linewidth]{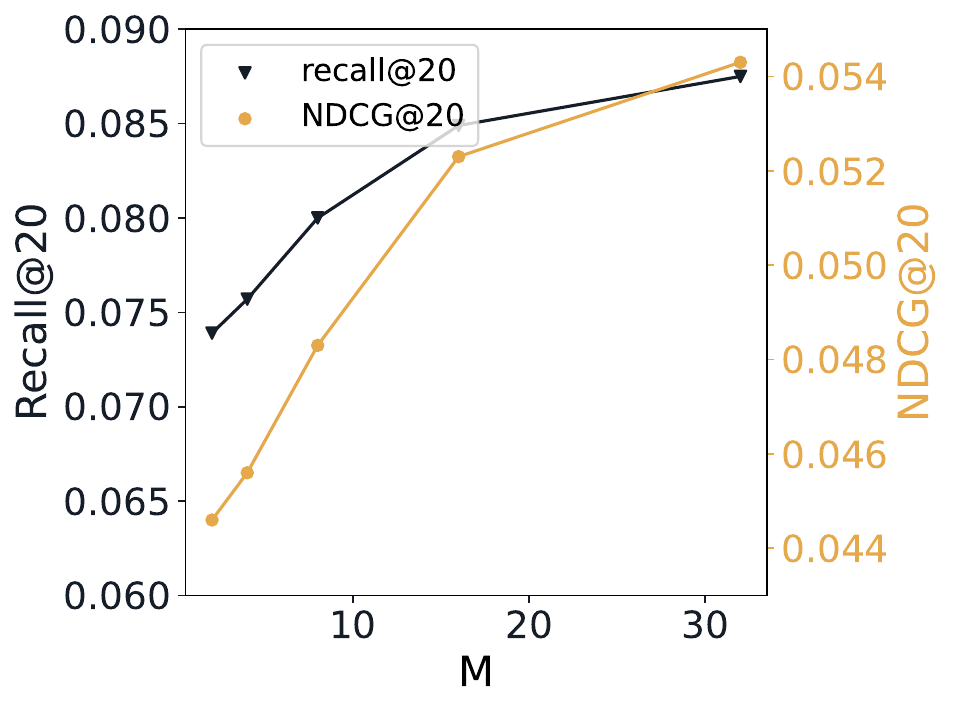}\end{minipage}} 
\subfigure[Gowalla]{\begin{minipage}[t]{0.38\linewidth}\centering\includegraphics[width=\linewidth]{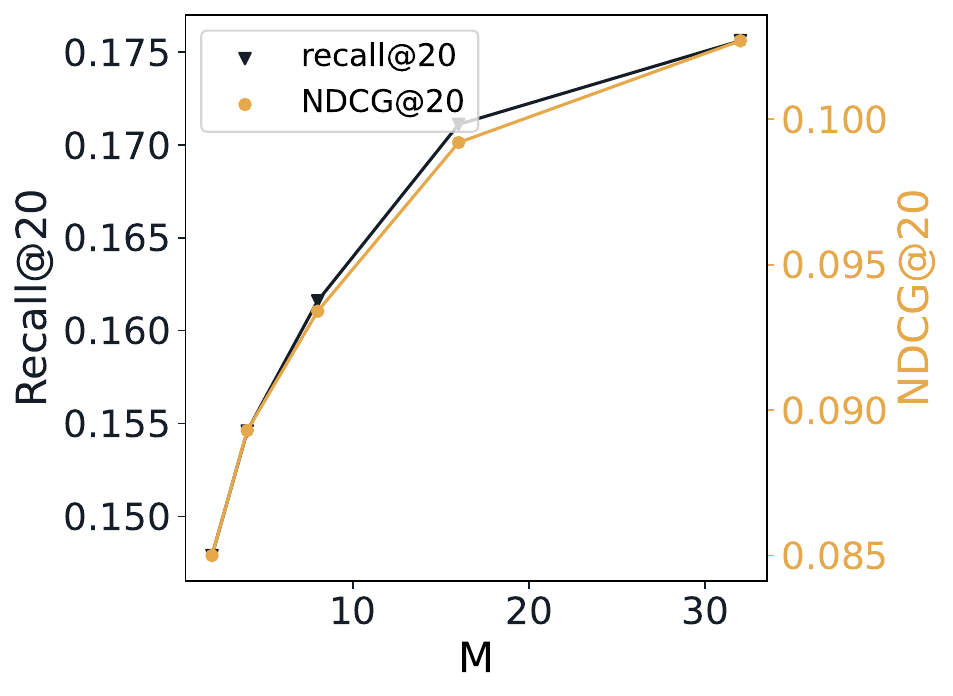}\end{minipage}} 
\vspace{-5mm}
\caption{Impact of $M$ in ANS.}
\label{fig:M}  
\vspace{-4mm}
\end{figure}
$M$ denotes the size of candidate negative set $\mathcal{E}$.
The present study illustrates the impact of $M$ on performance, as depicted in Figure~\ref{fig:M}. It is evident that an increase in $M$ leads to more negative samples, we can get harder negative samples by augmentation, thereby resulting in a performance improvement. Nevertheless, it is noteworthy that excessively large values of $M$ can considerably impede the efficiency of the models, and degrade performance due to noise (e.g., false negative samples).

\subsection{RQ6: Efficiency Analysis}
\begin{table}[t]
\begin{center}
\caption{Efficiency Analyses.}
\vspace{-3mm}
\begin{tabular}{@{}lcccc@{}}
\toprule
\multirow{1}{*}{\textbf{Dataset}} & \multirow{1}{*}{\textbf{Method}}   & \multirow{1}{*}{\textbf{training time}} & \multirow{1}{*}{\textbf{epoch}} \\ 
\midrule
\multirow{3}{*}{\textbf{ Amazon-Beauty}} 
& RNS & 0.47s & 95 \\
& DNS & 0.64s & 150  \\
& ANS & 0.89s & 145 \\

 \midrule
\multirow{3}{*}{\textbf{Gowalla}} 
& RNS & 0.23s & 90 \\
& DNS & 0.27s & 80  \\
& ANS & 0.43s & 140 \\

\bottomrule
\end{tabular}
\label{tab:Efficiency}
\vspace{-5mm}
\end{center}
\end{table}

Following the previous work~\cite{WYM22,MZW21}, we also examine the efficiency of different negative sampling methods in Table~\ref{tab:Efficiency}. We report the average running time (in seconds) per epoch and the total number of epochs needed before reaching convergence. All three methods are relatively efficient. There is no wonder that RNS is the most efficient method as it is a model-agnostic strategy. Compared to DNS, our proposed ANS method requires more running time. However, considering the huge performance improvement ANS brings, the additional running time is well justified.

\section{Conclusion}
Motivated by ambiguous trap and information discrimination, from which the state-of-the-art negative sampling methods suffer, for the first time, we proposed to introduce synthetic negative samples from a fine-granular perspective to improve implicit CF. We put forward a novel generic augmented negative sampling (ANS) paradigm, along with a concrete implementation. The paradigm consists of three major steps. The disentanglement step disentangles negative items into hard and easy factors in the absence of supervision signals. The augmentation step generates synthetic negative items using carefully calibrated noise. The sampling step makes use of a new metric called augmentation gain to effectively alleviate information discrimination. Comprehensive experiments demonstrate that ANS can significantly improve model performance and represents an exciting new research direction.

In our future work, we intend to explore the efficacy of augmented negative samples in tackling various issues such as fairness and popularity bias. Additionally, we will actively investigate the effectiveness of employing augmented negative sampling in online experiments.

\begin{acks}
This work was supported by the National Key R\&D Program of China (No.2020YFB1710200) and the National Natural Science Foundation of China (No.62072136).
\end{acks}
\balance
\bibliographystyle{ACM-Reference-Format}
\bibliography{recsys}

\clearpage
\appendix

\end{document}